# A Dataset of Weather Station-Scale Photosynthetic Phenology in Mid- to High-Latitude Regions of the Northern Hemisphere

Haiyang Shi[a,b*]

[a] State Key Laboratory of Desert and Oasis Ecology, Xinjiang Institute of Ecology and Geography,

Chinese Academy of Sciences, Urumqi 830011, China

[b] College of Resources and Environment, University of the Chinese Academy of Sciences, 19 (A) Yuquan

Road, Beijing, 100049, China.

*Corresponding authors: Haiyang Shi (Haiyang.shi@ms.xjb.ac.cn)

**Abstract**

This study extracted photosynthetic phenology indicators (SOS, POS, EOS) from daily GPP data at the weather-station scale for the mid- to high-latitude regions of the Northern Hemisphere from 2001 to 2019. The resulting dataset covers nearly 60,000 site-years, with more than 1,000 stations maintaining complete 19-year records. Results reveal pronounced spatial variability, with distinct trends across vegetation types and aridity zones. Compared with flux networks and satellite-based products, this dataset greatly enhances data availability for phenology studies and provides a solid foundation for assessing the impacts of local hydrothermal conditions and extreme climate events on phenology.

## 1 Introduction

In the context of global warming, phenological changes in the Northern Hemisphere have significant impacts on ecosystems and human society, as these changes directly affect key processes such as ecosystem productivity, carbon cycling, and water cycling (Li et al., 2023; Liu et al., 2016). Accurately predicting these changes not only helps in forecasting and managing ecosystem shifts, but also provides scientific guidance for agriculture and natural resource management, aiding society in better coping with the challenges posed by climate change.

Unlike traditional phenological events that focus on changes in plant structure (such as budding or leaf coloring), the seasonal cycle of photosynthesis is referred to as "vegetation photosynthetic phenology," which represents the functional aspect of plant activity (Fang et al., 2023). The start of the growing season (SOS), peak of the growing season (POS), and end of the growing season (EOS), all expressed as day of the year (DOY), are key indicators of photosynthetic phenology. SOS marks the beginning of active photosynthesis, POS represents the time when photosynthesis reaches its peak, and EOS signals the decline in plant activity:

1)   SOS changes are closely related to spring phenological events. An earlier spring leads to an earlier

SOS, potentially extending the growing season and increasing the ecosystem's capacity to absorb and store carbon. However, an earlier start might also deplete water and nutrient resources too soon, raising the risk of reduced carbon sequestration due to summer drought. Temperature is often considered the main driver of spring phenology changes in the Northern Hemisphere. Rising spring temperatures cause plants to bud and flower earlier because temperature directly affects plant physiological activities. Additionally, changes in daylight hours play an important role; longer daylight in spring enhances photosynthesis, promoting plant growth. Water availability is also crucial, as the timing and amount of spring precipitation influence soil moisture, which in turn affects plant growth and development.

2) POS changes are closely linked to the timing of carbon absorption by the ecosystem. An earlier spring can lead to an earlier POS, potentially lengthening the growing season and increasing carbon absorption. However, if POS occurs too early, plants may exhaust water and nutrient resources too quickly during the hot season, worsening summer drought and ultimately weakening the overall carbon absorption capacity of the growing season. Like SOS, POS is mainly driven by temperature, especially the rate of spring warming. An earlier POS might indicate that plants reach their photosynthetic peak sooner, reflecting a rapid response to early spring conditions. However, this early peak could also have negative effects, such as insufficient water availability later in the season, affecting plant growth and carbon absorption. Additionally, different types of vegetation may respond differently to changes in POS. For example, evergreen plants, which maintain photosynthesis year-round, show different POS patterns compared to deciduous plants.

3) EOS changes are closely related to autumn phenological events and mark the end of plant growth. In the context of climate warming, EOS may be delayed, which means a longer growing season, allowing plants more time to accumulate biomass and further increase carbon storage. However, a delayed EOS also comes with risks, such as increased exposure to late-season cold snaps, which could affect plant health and survival. Additionally, a delayed EOS may influence the dormancy period of plants and their growth in the following year. EOS is influenced not only by temperature but also by factors such as daylight duration, soil moisture, and nutrient availability.

The changes in SOS, POS, and EOS collectively impact the length and dynamics of the growing season, thereby affecting key ecosystem functions such as carbon cycling and water cycling. An extended growing season, particularly one caused by an earlier spring (earlier SOS) and a delayed autumn (later EOS), can significantly enhance the ecosystem's carbon absorption capacity. However, with rising temperatures and a longer growing season, ecosystems face increasing challenges in water resource utilization and nutrient cycling. The timing and height of the POS might also reflect the ecosystem's sensitivity to climate change. This sensitivity varies significantly across different vegetation types and geographic regions, further complicating the prediction and management of climate change impacts.

However, due to the limitations of various methods and data, the spatial and temporal characteristics and driving mechanisms of phenological changes in the Northern Hemisphere are still unclear. Large-scale studies of vegetation photosynthetic phenology often use satellite-based vegetation indices, such as NDVI and EVI, and more recently, satellite-based solar-induced chlorophyll fluorescence (SIF) to extract phenological indicators like SOS, POS, and EOS. While vegetation indices can partially represent photosynthesis and GPP, there can sometimes be a disconnect between the "greenness" indicated by these indices and actual photosynthetic activity. For example, these indices may struggle to capture photosynthetic changes in evergreen forests. Additionally, satellite-based vegetation indices have inherent limitations when representing the greenness of different vegetation types. The presence of snow cover in satellite images can also introduce uncertainties, especially when extracting spring phenology in high-latitude regions of the Northern Hemisphere. SIF data (Fang et al., 2023) has recently been widely used and is considered better at representing GPP compared to vegetation indices, but many studies have shown discrepancies between GPP indicated by SIF and that observed at flux towers. Furthermore, due to the relatively coarse resolution of SIF, large-scale phenological changes indicated by SIF are not easily validated at the site level.

Compared to flux towers, weather stations are more densely and widely distributed across the globe. Recent studies have shown that machine learning can estimate carbon and water flux data at the weather station scale with high accuracy (Shi et al., 2024). This suggests that we may be able to extract various phenological indicators at the weather station scale. Unlike SIF data, these station-scale data can be more easily validated, for example, by installing phenocams around weather stations. Previous studies often used reanalysis climate data to investigate the impact of meteorological factors on phenology. However, due to uncertainties in reanalysis data, they may not effectively represent the meteorological conditions that trigger phenological events. Reanalysis climate data (such as ERA5-land) is often considered effective in representing temperature but may have significant errors in representing water-related indicators such as vapor pressure deficit (VPD), the specific timing and amount of precipitation, frost dates, and the timing and intensity of extreme events. This poses a serious limitation for studying the factors influencing phenological changes, as accurate day-to-day data is crucial for this type of research. The coarse resolution of climate reanalysis data may lead to an underestimation of the impact of water-related factors (e.g., VPD, precipitation, soil moisture) on phenology. This is because sub-pixel spatial heterogeneity at a coarse resolution can cancel out finer-scale variations. Recent studies have shown that using observed data from weather stations, rather than reanalysis data, can better quantify the impact of VPD on carbon dynamics. This finding is likely important in studying the climate's influence on phenology as well. Temperature and moisture are key factors affecting phenology, and while coarse-resolution reanalysis data can capture temperature influences well at the regional scale where spatial heterogeneity is relatively small, it may not accurately reflect the impact of moisture due to its resolution limitations.Therefore, using observed meteorological data from stations can better reduce these errors, allowing for a more accurate capture of how meteorological factors influence phenology, at least at the station scale. Given that weather station records cover a much broader spatial and temporal range than flux towers, this also helps evaluate the impact of climate change on photosynthetic phenology over a wider area and longer timespan.

Therefore, this study produced a dataset capturing the changes in photosynthetic phenology (including annual SOS, POS, and EOS) across the Northern Hemisphere over the past two decades based on a daily GPP flux dataset estimated through machine learning at the weather station scale. This photosynthetic phenology dataset will complement existing large-scale, satellite-based phenology studies, allowing regions lacking flux towers to accurately identify SOS, POS, EOS and their driving mechanisms using weather station observations. This dataset will be of considerable significance for future studies on phenological changes in the context of climate change.

## 2 Methods

Daily GPP estimates at weather stations are from a developed GPP dataset (Untitled Item, 2024), which integrates the FLUXNET2015 dataset, the Global Surface Summary of the Day (GSOD) global weather station meteorological data, and satellite remote sensing inputs. The data development method is similar to previous studies on carbon flux estimation at weather stations. We applied a machine learning model, originally built using flux tower observations, to weather stations. Then, we assessed the expected accuracy of this model's predictions at the weather stations by evaluating the ecological and geographical similarity between each weather station and the flux towers. The underlying assumption is that a weather station located in an environment similar to that of a flux tower will have a similar level of modeling accuracy and predictability. Specifically, it is generated using a predictive model based on the Long Short-Term Memory (LSTM) network. GPP observations from FLUXNET2015 were used as the target variable. The model utilizes various predictor variables, including daily mean air temperature, maximum and minimum air temperatures, precipitation, wind speed, VPD, MODIS LAI, downward shortwave radiation , elevation, slope, and soil texture. The model demonstrates high accuracy, with most flux stations achieving an R-squared above 0.6 in a leave-one-station-out cross-validation. To assess the model's performance when applied to weather stations, we evaluated its extrapolative ability using the principle of geographic similarity. This approach assumes that stations with similar environmental conditions are likely to yield comparable prediction accuracies. Over 60% of weather stations achieved an R-squared greater than 0.5, with the highest accuracy observed predominantly in the Northern Hemisphere. For this study, we included only weather stations where the projected R-squared was above 0.5 to extract the SOS, POS and EOS information.

For each site year, only site years with a total number of days greater than 300 and with a number of days with data greater than 70 in spring (March, Apirl and May), summer (June, july and August), and autumn (September, October, and November) were retained in the extraction of phenology information. Middle and high latitudes in the Northern Hemisphere were set to have latitude values higher than 23.5. The daily GPP time series was smoothed using a Savitzky-Golay filter. This filter has been effectively employed in previous studies to reconstruct high-quality time series datasets, such as GPP (Xu et al., 2019) and the normalized difference vegetation index (Chen et al., 2004). The Savitzky-Golay filter coefficients were derived through an unweighted linear least-squares regression combined with a second-order polynomial model, and a 91-day span was chosen for the moving average to generate relatively smooth and stable daily GPP values (Xu et al., 2019). GPPmax was identified as the peak value in the smoothed GPP time series and POS was determined as the date when reaches GPPmax. SOS and EOS were determined using a fixed thre shold method, specifically the dates when the smoothed GPP reached 20% of the amplitude for the site (Huang et al., 2023; White et al., 2009). The amplitude was defined as the difference between GPPmax and the minimum of the smoothed GPP values. The length of the season was calculated as the difference between EOS and SOS and the time to peak was calculated as the difference between POS and SOS. Ultimately, a total of 57,829 site-years of SOS, POS, and EOS were extracted and included in the dataset.

In our analysis of the climate influences on SOS, POS, and EOS, we focused on water-related variables because moisture often shows higher variability and uncertainty in reanalysis climate data compared to temperature. We considered changes in VPD and precipitation during the preseason, as well as the 30 days and 7 days leading up to the events. We also included variables such as the time since the last rainfall and extreme VPD levels. For temperature, we used preseason, 30-day, and 7-day averages, and also considered the number of days since the last frost event (defined as a minimum temperature below 3°C), which may be crucial for SOS. Additionally, the previous year's EOS was included in the analysis of SOS, as past studies have shown that EOS from the previous year can significantly impact SOS. SOS was also used to analyze POS, as SOS and meteorological factors from SOS to POS can jointly influence POS. SOS and POS were considered in the analysis of EOS, as EOS may be affected by the carryover effects of SOS and POS, in addition to climate and other environmental factors.

## 3 Results

### 3.1 Data distribution

From 2001 to 2019, the number of available stations showed an upward trend, increasing from more than 2,000 in the early years to around 3,700 in the later years (Fig. 1). About 1,000 stations have complete data covering all 19 years, but many stations still have fewer than 10 years of records. Stations are more widely distributed in humid regions (Arid index > 0.65), which may be related to the dense coverage of stations in Europe and the United States. In terms of plant functional types (PFTs), cropland, grassland, and savannah account for the largest number of station-years. Forest station-years also exceed 2,000. Although the number of meteorological stations installed directly within forests is relatively small, the forest station-years in this dataset already far exceed those in flux observation networks such as FLUXNET2015.

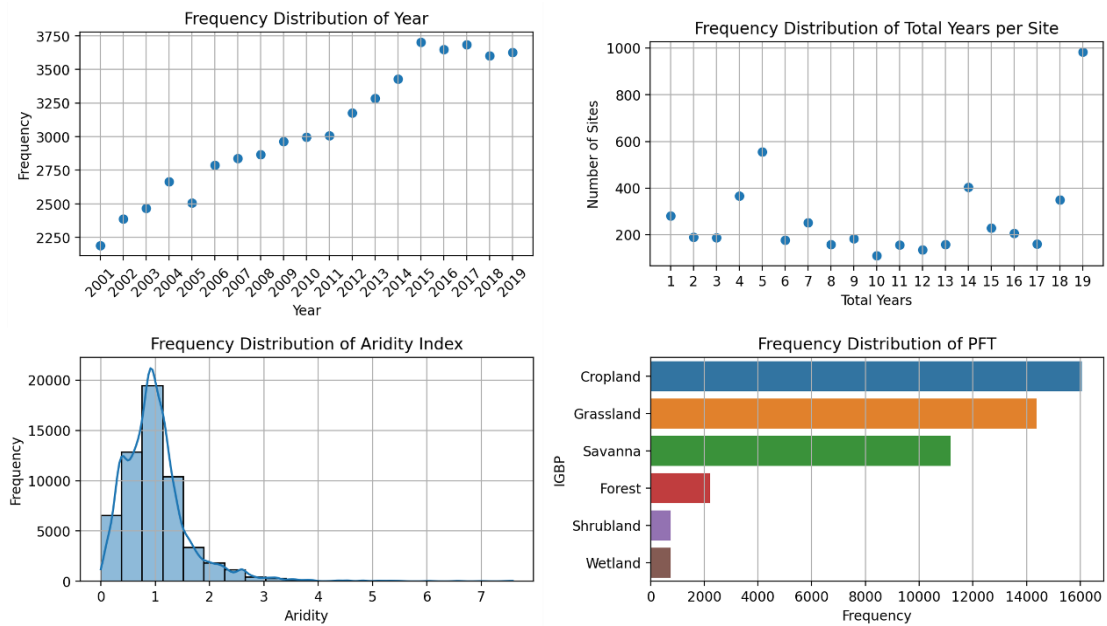

Figure 1. The number of sites per year, the distribution of the number of years per site, and the percentage of data in areas across aridity represented by the Arid Index (AI) and in different PFTs. The AI level classification (Middleton and Thomas, 1997): hyperarid (0 < AI < 0.05), arid (0.05 < AI < 0.2), semiarid (0.2 < AI < 0.5), and dry subhumid (0.5 < AI < 0.65).

**3.2 Spatial and temporal variations**

From 2001-2020 to 2011-2019, the spatial variability in phenological changes is quite pronounced (Fig. 2). SOS has advanced in the high-latitude regions of Eurasia, but it has been delayed in areas such as central and northeastern United States and southern China. Many sites in central U.S. are cropland, and although human activities (such as irrigation, fertilization, and other agricultural practices) may have a significant impact, growing degree days (GDD) and the timing of the last frost still largely dictate planting dates. The variability in POS is also considerable, with changes being more substantial than those observed for SOS and EOS. POS has generally advanced in the mid- to high-latitude regions of Eurasia, while in the southern U.S. and southern China, it has been delayed. The changes in POS and SOS show some synchronicity across Eurasia, but this is not as evident in North America. EOS exhibits even greater variability, with spatial patterns differing from those of SOS and POS. SOS has only advanced in shrublands (Fig. 3), while POS, conversely, has only been delayed in wetlands. EOS has been delayed in both shrublands and grasslands. In drylands, both SOS and POS have advanced, yet EOS has been delayed. In humid regions, SOS has been delayed, but both POS and EOS have advanced. These changes in SOS, POS, and EOS have also led to variations in GSL and TTP (Fig. 4). GSL has generally lengthened across much of Eurasia, especially in Europe, but has shortened in the central and eastern U.S. and southern China. TTP shows even greater spatial heterogeneity, with lengthening in the southern U.S., Western Europe, and southern China. Overall, the trends in GSL and TTP are not consistent, indicating significant variability in the duration from POS to EOS.

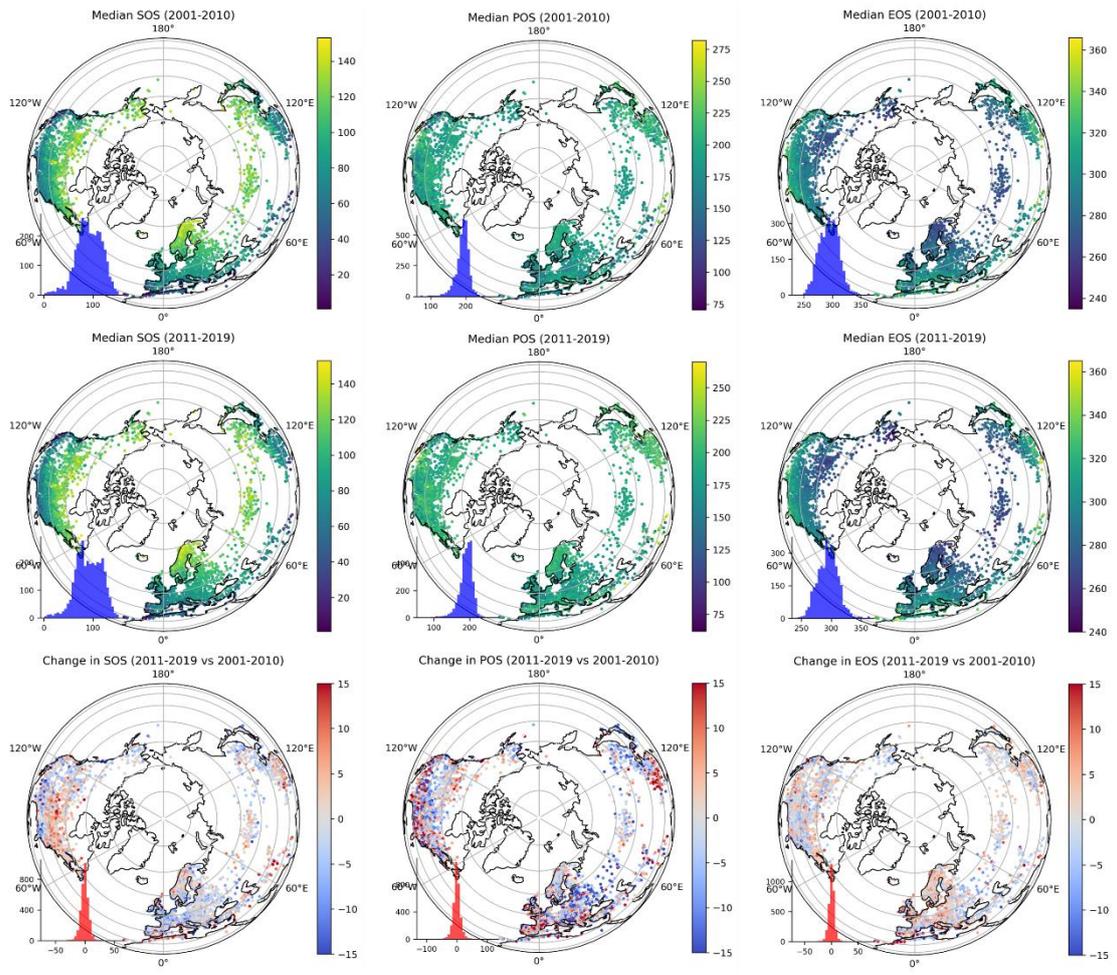

Figure 2. Median SOS, POS, EOS for two periods (2001-2010 and 2011-2019) and their changes. SOS, POS, and EOS are denoted using DOY values.

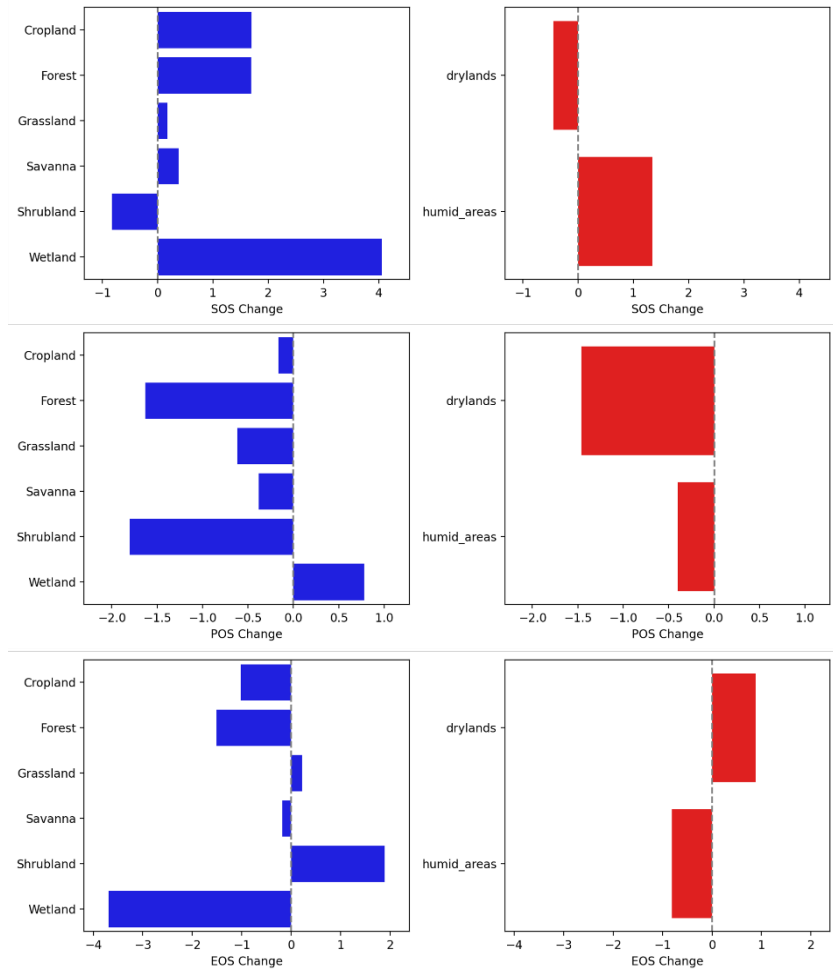

Figure 3. Mean values of changes in median SOS, POS, and EOS for different PFTs and different climate types (categorized into dryland and humid areas) for two periods (2001-2010 and 2011-2019).

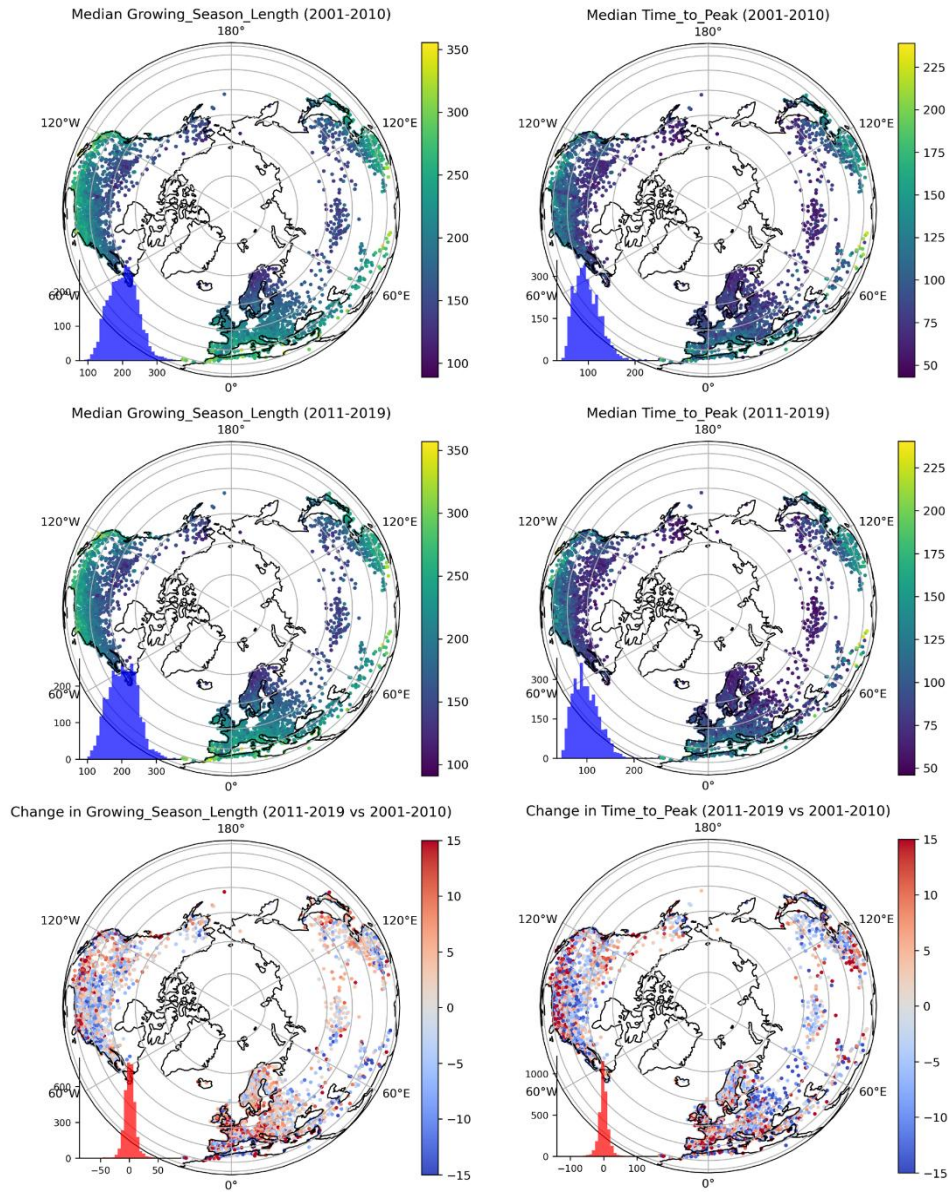

Figure 4. Median length of the growing season (EOS minus SOS) and time from the start of the growing season to the peak of the growing season (POS minus SOS) and their changes for two periods (2001-2010 and 2011-2019).

Overall, TTP and GSL show a significant positive correlation (r = 0.23, p < 0.05). However, considering that TTP is part of GSL, this correlation is not very strong. The duration from POS to EOS might weaken the relationship between GSL and TTP. SOS and EOS exhibit a significant negative correlation (r = -0.19, p < 0.05), indicating that warming leads to both a delayed EOS and an advanced SOS, effectively extending the growing season. The correlation between POS and SOS is lower (r = 0.05), though still significant. The correlation between changes in EOS and POS is not significant, likely due to the more complex mechanisms influencing EOS.

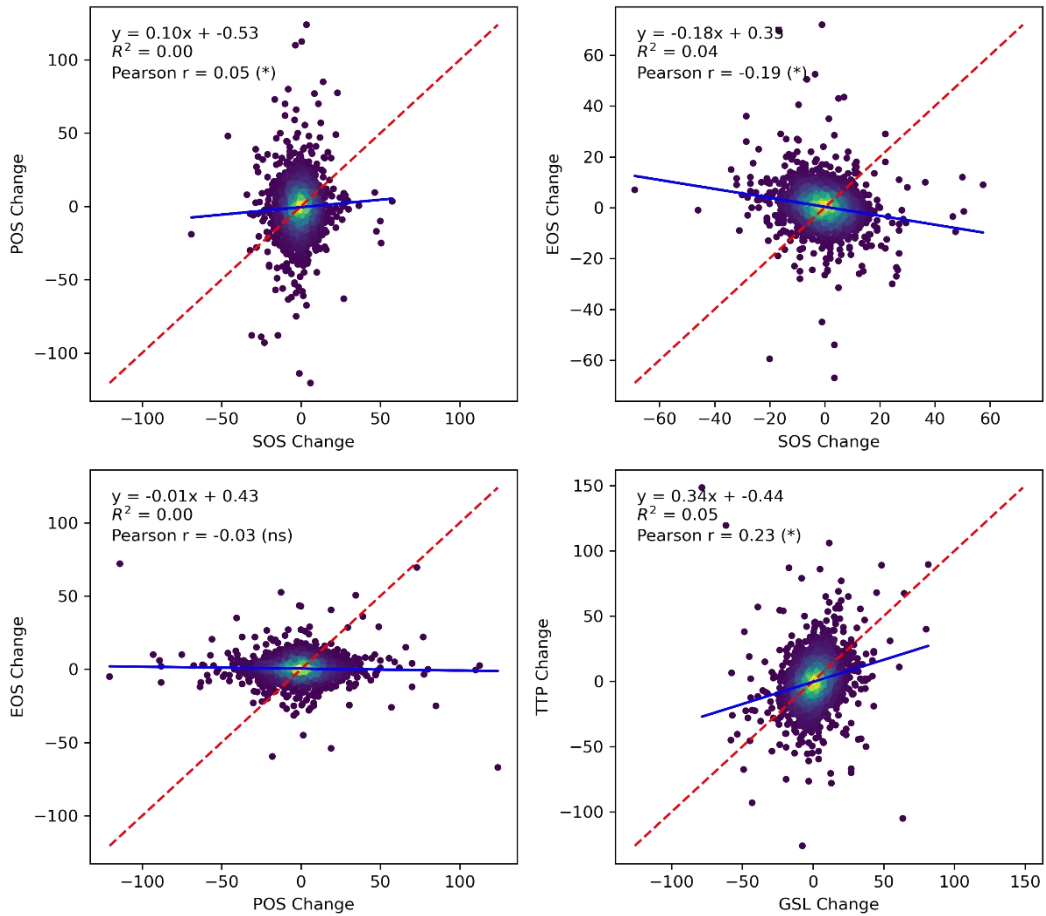

Figure 5. The relationships among the changes in SOS, POS, and EOS, and the relationships between changes in GSL and TTP.

**4 Discussions**

The weather station–based photosynthetic phenology dataset developed in this study (including SOS, POS, and EOS) substantially expands the data foundation available for phenology research. Compared with traditional approaches that rely on satellite vegetation indices (e.g., NDVI, EVI, SIF) or flux tower observations, this dataset offers several clear advantages. First, weather stations are more widely distributed, particularly in humid mid- to high-latitude regions of Europe and North America, thereby providing broader coverage and enabling more comprehensive assessments of regional phenological patterns and trends. Second, the dataset contains more complete and continuous time series, with more than 1,000 stations having full 19-year records, which greatly enhances the ability to assess long-term dynamics. Third, it encompasses a wide range of plant functional types (PFTs), including croplands, grasslands, and savannahs, with the number of forest station-years exceeding that of existing flux networks such as FLUXNET2015, thereby offering valuable resources for forest phenology research.

Beyond enhancing data availability, this dataset provides unique opportunities to evaluate the influence of local hydrothermal conditions on photosynthetic phenology. Unlike coarse-resolution reanalysis products, station-based observations can better capture day-to-day variations in water-related variables such as precipitation timing, soil moisture conditions, and vapor pressure deficit (VPD), which are critical drivers of phenological events. In addition, the dataset enables the investigation of the impacts of extreme climate events on phenology. For example, spring frosts, summer droughts, or autumn heat waves may significantly shift the timing of SOS, POS, and EOS. With this dataset, it becomes possible to quantify how such events alter the length of the growing season (GSL) and the time to peak (TTP), thereby advancing our understanding of the vulnerability of carbon and water cycles under climate change.

In summary, the weather station–based photosynthetic phenology dataset developed here not only complements existing satellite- and tower-based studies but also provides a solid foundation for future research on climate impacts, ecosystem carbon sink potential, and agricultural management in the context of global change.

**5 Conclusions**

The weather-station-based photosynthetic phenology dataset developed in this study substantially enriches the spatial and temporal resources available for phenology research, with advantages in coverage, temporal continuity, and vegetation type diversity. It not only supports the detection of long-term trends in photosynthetic phenology under climate warming but also enables the evaluation of local hydrothermal influences and the impacts of extreme climate events. Looking ahead, this dataset provides an important foundation for carbon cycle modeling, ecosystem vulnerability assessments, and agricultural management.

## Author Contributions

HS initiated this research and were responsible for the integrity of the work as a whole. HS performed formal analysis and calculations of SOS, EOS and POS. HS drafted the manuscript.

## Data availability

The extracted SOS, POS and EOS dataset (Shi, 2024) is available at https://doi.org/10.5281/zenodo.13292972